\magnification=1150
\parskip=10pt
\parindent=14pt
\baselineskip=15pt
\pageno=0
\footline={\ifnum \pageno <1 \else \hss \folio \hss \fi}
\topglue 1.2in
\centerline{\bf The Quantum Cosmological Wavefunction at Very Early Times}
\centerline{\bf for a Quadratic Gravity Theory}
\vskip .5in
\centerline{\bf Simon Davis}
\vskip .5in
\centerline{\bf Institut f{\"u}r Mathematik}
\centerline{\bf Universit{\"a}t Potsdam}
\centerline{\bf D-14415 Potsdam}
\vskip .5in
\noindent
{\bf Abstract.} The quantum cosmological wavefunction for a quadratic gravity
theory derived from the heterotic string effective action is obtained
near the inflationary epoch and during the initial Planck era.  Neglecting
derivatives with respect to the scalar field, the wavefunction would satisfy
a third-order differential equation near the inflationary epoch which has 
a solution that is singular in the scale factor limit $a(t)\to 0$.
When scalar field derivatives are included, a sixth-order differential
equation is obtained for the wavefunction and the solution by Mellin transform
is regular in the $a\to 0$ limit.  It follows that inclusion of the scalar 
field in the quadratic gravity action is necessary for consistency of the 
quantum cosmology of the theory at very early times.

\vfill\eject

The quadratic gravity theory derived from the four-dimensional heterotic 
string effective action obtained by orbifold compactification of the extra
dimensions contains the coupling of the dilaton field to the 
Gauss-Bonnet term $R_{GB}^2$ and a pseudo-scalar field to an $R{\tilde R}$ 
term [1].  While the quantum cosmology of this theory can be defined on a 
superspace which includes metrics with anisotropy, the properties of 
renormalizability in the generalized sense and unitarity occur only when 
the $R{\tilde R}$ term is set equal to zero.  This can be achieved either 
by setting the pseudo-scalar field equal to zero without necessarily imposing 
any restrictions on the metric superspace or by confining the quantum 
cosmology to a minisuperspace of isotropic metrics.  If the minisuperspace 
is chosen to be the space of Friedmann-Robertson-Walker metrics with scale 
factors $\{a(t)\}$ and scalar fields $\{\Phi(t)\}$, integration over the 
spatial coordinates gives a volume factor and leaves a one-dimensional action 
with an ${\ddot a}{\dot a}^2$ term.  While a boundary term can be added to 
produce an action depending only on $a, {\dot a}$ and $\Phi$, different 
quantization procedures may be used according to the choice of conjugate 
momentum variables.  For an action with the second derivative of the scale 
factor, the conjugate momentum $P_{\dot a}$ can be introduced leading to a 
closed-form differential equation for the quantum cosmological wavefunction 
[2].  Use of the conventional momenta $P_a,~P_\Phi$ gives rise to a Wheeler-De
Witt equation which can be expanded in powers of ${{e^{-\Phi}}\over {g_4^2}}$  
with the higher-order terms representing corrections to the second-order
Wheeler-DeWitt equation [3].  At very early times, however, the higher-order
curvature terms will have a magnitude comparable to the Ricci scalar, so 
that a truncation of the series expansion in ${{e^{-\Phi}}\over {g_4^2}}$
would no longer accurately define the Hamiltonian.  Consequently, the
sixth-order closed-form differential equation is preferable for the initial
Planck era, whereas the series expansion of the Hamiltonian in
${{e^{-\Phi}}\over {g_4^2}}$ would be valid in the inflationary
epoch.  While this procedure might be considered to be at variance with the
standard canonical quantization of a gravitational field theory, different 
methods of quantization also have been found to be necessary to obtain the 
appropriate form for the Wheeler-DeWitt equation for higher-order gravity [4].
It also would be consistent with the freezing of the additional degree of 
freedom $P_{\dot a}$ during the transition from the Planck era to the 
inflationary epoch.  For an inflationary model based on a slow-roll scalar 
potential, derivatives of the wavefunction with resepect to $\Phi$ can
be neglected [5].  This approximation can be used to obtain a limiting form 
of the closed-form differential equation near the boundary between the Planck
era and the inflationary epoch, which is a third-order differential 
equation in the scale factor $a$.  After imposing matching conditions 
on the wavefunction and its first two derivatives at the boundary, the 
numerical solution extrapolated to $a=0$ is found to be singular.  An exact 
solution of the sixth-order differential equation by series expansion 
can be obtained by deducing the coefficients from a two-variable 
recursion relation.  Techniques have recently been developed for 
solving two-variable recursion relations [6], but the approximate solution 
by Mellin transform is sufficient to establish regularity of the wavefunction 
in the $a\to 0$ limit.  

\noindent
{\bf 1. Solution to Differential Equation for the Wavefunction near the
Inflationary}
\hfil\break
\phantom{......}{\bf Epoch}

Given an isotropic cosmology and specializing to Friedmann-Robertson-Walker
metrics, the low-energy effective action for the heterotic string can be
reduced to a one-dimensional integral 
$$I~=~\int~dt~\left[(6a^2{\ddot a}+6a{\dot a}^2+6aK)+{1\over 2}a^3{\dot \Phi}^2
+6{{e^{-\Phi}}\over {g_4^2}}{\ddot a}({\dot a}^2+K)\right]
\eqno(1)
$$
with conjugate momenta [2][3]
$$\eqalign{P_a~&=~{{\partial L}\over {\partial {\dot a}}}-{d\over {dt}}
\left({{\partial L}\over {\partial {\ddot a}}}\right)
~=~6 {{e^{-\Phi}}\over {g_4^2}}{\dot \Phi}({\dot a}^2+K)
\cr
P_{\dot a}~&=~6\left(a^2+{{e^{-\Phi}}\over {g_4^2}}({\dot a}^2+K)\right)
\cr
P_\Phi~&=~{{\partial L}\over {\partial {\dot \Phi}}}~=~a^3{\dot \Phi}
\cr}
\eqno(2)
$$
so that the Hamiltonian is
$$\eqalign{H~&=~P_a {\dot a}+P_{\dot a} {\ddot a}+P_\Phi {\dot \Phi}-L
\cr
&= -6a({\dot a}^2+K)+6{{e^{-\Phi}}\over {g_4^2}}{\dot \Phi}({\dot a}^2+K)
{\dot a}+{1\over 2}a^3{\dot \Phi}^2
\cr
&=-g_4^2P_\Phi^{-1}e^\Phi a^4 P_a^4~+~{1\over {2a^3}}P_a^2~+~
\left[{{g_4^2}\over 6}P_\Phi^{-1}e^\Phi P_a^2 a^3 P_a~-~KP_a^2
\right]^{1\over 2}
\cr}
\eqno(3)
$$
and the Wheeler-DeWitt equation $H\Psi=0$ can be transformed into
$$\eqalign{-{{g_4^2}\over 6}e^{-\Phi}
&\left(a^3{{\partial^4\Psi}\over {\partial a^3
\partial \Phi}}+6a^2 {{\partial^3\Psi}\over {\partial a^2 \partial \Phi}}
\right)
+Ke^{-2\Phi}\left({{\partial^4 \Psi}\over {\partial a^2 \partial \Phi^2}}
-{{\partial^3 \Psi}\over {\partial a^2 \partial \Phi}}\right)
\cr
~&=~a^4g_4^4\left[4a^3 {{\partial \Psi}\over {\partial a}}+ a^4
{{\partial^2\Psi}\over {\partial a^2}}\right]~+~ag_4^2e^{-\Phi}\left(
{{\partial^4\Psi}\over {\partial a \partial \Phi^3}}+{{\partial^3\Psi}\over
{\partial a \partial \Phi^2}}\right)
\cr
&~~~~~~~~~~~~~~+~{3\over 2}g_4^2 e^{-\Phi}
\left(a{{\partial^2\Psi}\over {\partial a \partial \Phi}}-
{{\partial^3\Psi}\over {\partial \Phi^3}}\right)~+~
{1\over {4a^6}}e^{-\Phi}{\partial\over {\partial \Phi}}
\left(e^{-\Phi}{{\partial^5\Psi}\over {\partial \Phi^5}}\right)
\cr}
\eqno(4)
$$
The correspondence between the solutions of the sixth-order equation and
the pseudo-
\hfil\break
differential equation $H\Psi=0$ can be defined as follows.
Given the two operators
$$\eqalign{{\cal A}~&=~g_4^2P_\Phi^{-1}e^\Phi a^4 P_a
                          ~-~{1\over {2a^3}}P_\Phi^2
\cr
{\cal B}~&=~\left[{{g_4^2}\over 6}P_\Phi^{-1}e^\Phi P_a^2 a^3 P_a~-~KP_a^2
\right]^{1\over 2}
\cr}
\eqno(5)
$$
the solutions to $H\Psi=0$ belong to $ker~({\cal A}-{\cal B})$.  If
${\cal C}=e^{-\Phi}P_\Phi e^{-\Phi}P_\Phi ({\cal A}^2-{\cal B}^2)$
then $ker~{\cal C}=ker({\cal A}^2-{\cal B}^2)\cup ({\cal A}^2-
{\cal B}^2)^{-1}~ker(e^{-\Phi}P_\Phi e^{-\Phi}P_\Phi)$.  Since 
$({\cal A}^2-{\cal B}^2)\Psi=({\cal A}+{\cal B})({\cal A}-{\cal B})\Psi
~+~({\cal A}{\cal B}-{\cal B}{\cal A})\Psi$, 
$$\eqalign{[ker({\cal A}-{\cal B})&~\cup~({\cal A}-{\cal B})^{-1}
                   ker({\cal A}+{\cal B})]
                  ~\cap~\ker({\cal A}{\cal B}-{\cal B}{\cal A})
                    ~\subset~ker({\cal A}^2-{\cal B}^2)
\cr
ker({\cal A}-{\cal B})&=ker[({\cal A}+{\cal B})({\cal A}-{\cal B})]
\cap ({\cal A}-{\cal B})^{-1}~coker({\cal A}+{\cal B})
\cr             
&\supset[ker({\cal A}^2-{\cal B}^2)
                 ~\cap~\ker({\cal A}{\cal B}-{\cal B}{\cal A})]
                  \cap  ({\cal A}-{\cal B})^{-1}~coker({\cal A}+{\cal B})
\cr}    
\eqno(6)        
$$
If $\Psi\in ker({\cal A}-{\cal B})$, then 
$({\cal A}{\cal B}-{\cal B}{\cal A})\Psi=({\cal A}^2-{\cal B}^2)\Psi$,
so that the equivalence
$ker({\cal A}-{\cal B})=ker({\cal A}^2-{\cal B}^2)\cap 
                        ({\cal A}-{\cal B})^{-1}~coker({\cal A}+{\cal B})
$
is valid in this space.   As $ker({\cal A}^2-{\cal B}^2)= 
{\cal A}^{-1}~ker({\cal A}-{\cal B})\cap {\cal B}^{-1}~ker({\cal A}-{\cal B})$,
the restriction to the subdomain $ker({\cal A}-{\cal B})$ implies that
$ker({\cal A}^2-{\cal B}^2)\vert_{ker({\cal A}-{\cal B})}
={\cal A}^{-1}~ker({\cal A}-{\cal B})$ 
since ${\cal A}^{-1}\Psi={\cal B}^{-1}\Psi$.
Because
$$ker({\cal A}^2-{\cal B}^2)=ker~{\cal C}\cap ({\cal A}^2-{\cal B}^2)^{-1}
                                   coker(e^{-\Phi}P_\Phi e^{-\Phi}P_\Phi)
\eqno(7)
$$
the subdomain of $ker~{\cal C}$ which satisfies ${\cal A}\Psi={\cal B}\Psi$
is given by
$$\biggl[ker~{\cal C}\cap ({\cal A}^2-{\cal B}^2)^{-1}
                                   coker(e^{-\Phi}P_\Phi e^{-\Phi}P_\Phi)
                               \biggr]\bigg\vert_{ker({\cal A}-{\cal B})}
                        \cap ({\cal A}-{\cal B})^{-1}~coker({\cal A}+{\cal B}) 
                           \subset ker({\cal A}-{\cal B})
\eqno(8)
$$
However, the spaces of functions $coker(e^{-\Phi}P_\Phi e^{-\Phi}P_\Phi)$ and
$coker({\cal A}+{\cal B})$ are sufficiently large that the intersection
does not exclude a significant portion of $ker~{\cal C}\vert_{ker~({\cal A}
-{\cal B})}$.  

With the addition of the potential term $V(\Phi)$, the Hamiltonian becomes
$$H~=~-g_4^2P_\Phi^{-1} e^\Phi a^4P_a~+~{1\over {2a^3}}P_\Phi^2
~+~\left[{{g_4^2}\over 6}P_\Phi^{-1}e^\Phi P_a^2 a^3 P_a-KP_a^2
\right]^{1\over 2}~+~a^3V(\Phi)
\eqno(9)
$$
Transposing the square root expression leads to the equation
$$\left(g_4^2P_\Phi^{-1}e^\Phi a^4 P_a~-~{1\over {2a^3}} P_\Phi^2~-~a^3V(\Phi)
\right)\Psi~=~\left[{{g_4^2}\over 6}P_\Phi^{-1}e^\Phi P_a^2 a^3 P_a
-KP_a^2\right]^{1\over 2}\Psi
\eqno(10)
$$
and upon squaring this equation, the new terms are
$$\eqalign{\left(g_4^2P_\Phi^{-1}e^\Phi a^4 P_a
-{1\over {2a^3}}P_\Phi^2\right)&(-a^3V(\Phi))\Psi
\cr
&~~~~+~(-a^3V(\Phi))
\biggl(g_4^2P_\Phi^{-1}e^\Phi a^4 P_a~-{1\over {2a^3}}P_\Phi^2\biggr)\Psi
~+~a^6 V(\Phi)^2\Psi
\cr}
\eqno(11)
$$
Multiplying by a single power of $e^{-\Phi}P_\Phi$ gives
$$\eqalign{\biggl(g_4^2 a^4 P_a a^4 P_\Phi^{-1} e^\Phi
&~-~{{g_4^2}\over 2} P_\Phi^2 e^\Phi a P_a
~+~{1\over {4a^6}}e^{-\Phi}P_\Phi^5~-~{{g_4^2}\over 2} a^4 P_a 
{1\over {a^3}} P_\Phi^2\biggr)\Psi
\cr
&~=~{{g_4^2}\over 6}P_a^2 a^3 P_a \Psi~-~Ke^{-\Phi}P_\Phi P_a^2 \Psi
\cr}
\eqno(12)
$$
Since $P_\Phi$ is represented by the differential operator 
$-i{\partial\over {\partial \Phi}}$ and its inverse $P_\Phi^{-1}$ by 
the integral $i\int~d\Phi$.  For a slow-roll potential $\vert V^{-1}(\Phi)
V^\prime(\Phi)\vert\ll 1$ and given that the $\Phi$ derivative of the potential
and wave function is negligible in the semi-classical regime, 
$$-ig_4^2 a^4 {\partial\over {\partial a}} a^4 {\partial\over {\partial a}}
\int~d\Phi~e^\Phi \Psi+2ig_4^2 a^7 V(\Phi) {{\partial \Psi}\over 
{\partial a}}+3ig_4^2 a^6 V(\Phi) \Psi=i{{g_4^2}\over 6}
{{\partial^2}\over {\partial a^2}}\left(a^3{{\partial \Psi}\over {\partial a}}
\right)
\eqno(13)
$$
giving rise to a third-order differential equation
$${{a^2}\over 6} {{d^3\Psi}\over {da^3}}~+~a(1+e^\Phi a^6){{d^2\Psi}\over 
{da^2}}~+~[(1+e^\Phi a^6)-2a^6 V(\Phi)]{{d\Psi}\over {da}}
~-~3a^5 V(\Phi)\Psi~=~0
\eqno(14)
$$
The solution to this equation can be matched with the wave function satisfying
the second Wheeler-DeWitt equation in the inflationary epoch after
specification of the boundary conditions.   This equation differs from
a standard equation with form [7]
$$x^2 y^{\prime\prime\prime}+x(ax^n+b+c+1)y^{\prime\prime}+[\alpha x^{2n}+(ac+
\beta)+\gamma+bc]y^\prime+(c-1)(\alpha x^{2n}+\beta x^n+\gamma){y\over x}=0
\eqno(15)
$$
by a single term, after setting $\gamma=0$, $\alpha=0$,
$n=6$, $a=6e^\Phi$, $b+c+1=6$, $bc=6$, $ac +\beta=12e^\Phi+\beta
=6e^\Phi-12V(\Phi)$. It is known that the substitution $w=xy^\prime
+(c-1)y$ leads to a second-order equation of the form
$x^2w^{\prime\prime}+x(ax^n+b)w^\prime+(\alpha x^{2n}+\beta x^n+\gamma)y=0$.

The $a\to 0$ limit of the differential equation (10) is
$${{a^2}\over 6} {{d^3\Psi}\over {da^3}}+a{{d^2\Psi}\over {da^2}}+
{{d\Psi}\over {da}}=0
\eqno(16)
$$
or equivalently
$$a^2{{d^2\chi}\over {da^2}}+a{{d\chi}\over {da}}+6\chi=0
\eqno(17)
$$
with ${{d\Psi}\over {da}}$ set equal to $\chi$.  If
$$H_0~=~a^2 {{d^2}\over {da^2}}~+~6a{d\over {da}}~+~6
\eqno(18)
$$
the full equation is 
$$\eqalign{(H_0+H_1)\chi&=0
\cr
H_1\chi&=6a^7e^\Phi{{d\chi}\over {da}}+6a^6(e^\Phi-2V(\Phi))\chi
                   -18a^5V(\Phi)\int da \chi
\cr}
\eqno(19)
$$
Since the solution to $H_0\chi_0=0$ is $\chi_0={A\over {a^2}}+{B\over {a^3}}$,
$$\eqalign{
H_1\chi_0&=6a^7e^\Phi\left(-{{2A}\over {a^3}}-{{3B}\over {a^4}}\right)
               +6a^6(e^\Phi-2V(\Phi))\left({A\over {a^2}}+{B\over {a^3}}\right)
               +18a^5V(\Phi)\left({A\over a}+{B\over {2a^2}}-C\right)
\cr
&=6Aa^4(V(\Phi)-e^\Phi)-3B(V(\Phi)+4e^\Phi)-18Ca^5 V(\Phi)
\cr}
\eqno(20)
$$
If $H_0\chi_1=-H_1\chi_0$, then $H(\chi_0+\chi_1)=H_1\chi_1$.  With the
additional terms, a series solution $\sum_{n=0}^\infty \chi_n$ is obtained
for $\chi$ with $H_0\chi_{n+1}=-H_1\chi_1$.  Defining the Wronskian to be
$$W=\chi_{01}{d\over {da}}\chi_{02}-\chi_{02}{d\over {da}}\chi_{01}
={A\over {a^2}}{d\over {da}}{B\over {a^3}}-{B\over {a^3}}{d\over {da}}
{A\over {a^2}}=-{{AB}\over {a^6}}
\eqno(21)
$$
and noting that the differential operator $H_0$ is $a^{-4}{d\over {da}}
a^6 {d\over {da}}+6$, the solution to the equation $H_0\chi_1=-H_1\chi_0$
is
$$\chi_1=k_1\chi_{01}+k_2\chi_{02}+\chi_{02}\int da a^{-2} \chi_{01}
{{(-H_1\chi_0)}\over W}-\chi_{01}\int da a^{-2}\chi_{02}{{(-H_1 \chi_0)}\over
W}
$$
which equals
$$\eqalign{\chi_1&={A\over {a^2}}\int da {B\over {a^3}} a^{-2}
{{[6Aa^4(V(\Phi)-e^\Phi)-3Ba^3(V(\Phi)+4e^\Phi)+18Ca^5V(\Phi)]}
\over {{AB}\over {a^6}}}
\cr
&~~~~-{B\over {a^3}}\int da {A\over {a^2}} a^{-2}
{{[6Aa^4(V(\Phi)-e^\Phi)-3Ba^3(V(\Phi)+4e^\Phi)+18Ca^5V(\Phi)]}
\over {{AB}\over {a^6}}}
\cr
&={B\over {10}}a^3(V(\Phi)+4e^\Phi)-{A\over 7}a^4(V(\Phi)-e^\Phi)
+{{9C}\over {28}}a^5V(\Phi)
\cr}
\eqno(23)
$$
after setting $k_1=k_2=0$, since an additional constant in the expression for
each integral only leads to an overall shift in the value of $A$ and $B$ 
and therefore can be absorbed into those parameters.  Then
$$\chi_0+\chi_1={A\over {a^2}}+{B\over {a^3}}+{B\over {10}}a^3(V(\Phi)+4e^\Phi)
-{A\over 7}a^4(V(\Phi)-e^\Phi)+{{9C}\over {28}}a^5V(\Phi)
\eqno(24)
$$
and
$$\Psi_0+\Psi_1=-{A\over a}-{B\over {2a^2}}+D+{B\over {40}}a^4(V(\Phi)+
4e^\Phi)-{A\over {35}}a^5(V(\Phi)-e^\Phi)-{{3C}\over {56}}a^6V(\Phi)
\eqno(25)
$$
Substitution of this function into the third-order differential equation gives
$D=0$, except at $a=0$.  Then
$$\Psi_0+\Psi_1=-{A\over a}-{B\over {2a^2}}+{B\over {40}}(V(\Phi)+4e^\Phi)a^4
-{A\over {35}}(V(\Phi)-e^\Phi)a^5-{{3C}\over {56}}V(\Phi)a^6~~~~~~a\ne 0
\eqno(26)
$$
At $a=0$, it is possible to set $D=1$ to satisfy the initial condition 
$\Psi(a=0)=1$, although $D=0$ for all $a$ would be necessary for continuity
of the wavefunction.

At the next order
$$\eqalign{H_1&\left[-{{a^4}\over 7}(V(\Phi)-e^\Phi)A+{{a^3}\over {10}}
(V(\Phi)+4e^\Phi)B+{{9C}\over {28}}a^5V(\Phi)\right]
\cr
&=6a^7e^\Phi\left[-{{4a^3}\over 7}(V(\Phi)-e^\Phi)A+{{3a^2}\over {10}}
(V(\Phi)+4e^\Phi)B+{{45C}\over {28}}a^4V(\Phi)\right]
\cr
&~~~+~6a^6(e^\Phi-2V(\Phi))\left[-{{a^4}\over 7}(V(\Phi)-e^\Phi)A+
{{a^3}\over {10}}(V(\Phi)+4e^\Phi)B+{{9C}\over {28}}a^5V(\Phi)\right]
\cr
&~~~-~18a^5V(\Phi)\left[-{{a^5}\over {35}}(V(\Phi)-e^\Phi)A+
{{a^4}\over {40}}(V(\Phi)+4e^\Phi)B+{{3C}\over {56}}a^6V(\Phi)
\right]
\cr
&=A{{-150e^\Phi+78V(\Phi)}\over {35}}a^{10}(V(\Phi)-e^\Phi)
+B{{48e^\Phi-33V(\Phi)}\over {20}}a^9(V(\Phi)+4e^\Phi)
\cr
&~~~+C\left({{81}\over 7}e^\Phi-{{135}\over {28}}V(\Phi)\right)
a^{11}V(\Phi)
\cr}
\eqno(27)
$$
and the solution to the equation $H_0\chi_2=-H_1\chi_1$ is
$$\eqalign{
\chi_2&=-{1\over {a^2}}\int da~a \biggl[A{{-150e^\Phi+78V(\Phi)}\over {35}}
a^{10}(V(\Phi)-e^\Phi)+B{{48e^\Phi-33V(\Phi)}\over {20}}a^9(V(\Phi)+4e^\Phi)
\cr
&~~~~~~~~~~~~~~~~~~~~~~~~+~C\left({{81}\over 7}e^\Phi-{{135}\over {28}}
V(\Phi)\right)a^{11}V(\Phi)\biggr]
\cr
&~~+{1\over {a^3}}\int da~a^2 \biggl[A{{-150e^\Phi+78V(\Phi)}\over {35}}a^{10}
                                         (V(\Phi)-e^\Phi)
+B{{48e^\Phi-33V(\Phi)}\over {20}}a^9(V(\Phi)+4e^\Phi)
\cr
&~~~~~~~~~~~~~~~~~~~~~~~~~~+~C\left({{81}\over 7}e^\Phi-{{135}\over {28}}
V(\Phi)\right)a^{11}V(\Phi)\biggr]
\cr
&=A{{(75e^\Phi-39V(\Phi))(V(\Phi)-e^\Phi)a^{10}}\over {2730}}
-B{{(16e^\Phi-11V(\Phi))(V(\Phi)+4e^\Phi)a^9}\over {880}}
\cr
&~~~~~~~~~~~~~~~~~~~~~~~~~~~~~~~
-C{1\over {182}}\left({{81}\over 7}e^\Phi-{{135}\over {28}}\right)V(\Phi)a^{11}
+...
\cr}
\eqno(28)
$$
It follows that
$$\eqalign{
\chi&={A\over {a^2}}+{B\over {a^3}}+{B\over {10}}(V(\Phi)+4e^\Phi)a^3
-{A\over 7}(V(\Phi)-e^\Phi)a^4+{{9C}\over {28}}V(\Phi)a^5
\cr
&-{B\over {880}}(16e^\Phi-11V(\Phi))(V(\Phi)+4e^\Phi)a^9
+{A\over {2730}}(75e^\Phi-39V(\Phi))(V(\Phi)-e^\Phi)a^{10}+
\cr
&~~~~~~~~~~~~~~~~~~~~~~~~~~~~~~~~
-C{1\over {182}}\left({{81}\over 7}e^\Phi-{{135}\over {28}}\right)V(\Phi)a^{11}
+...
\cr}
\eqno(29)
$$
and
$$\eqalign{\Psi&=-{A\over a}-{B\over {2a^2}}+{B\over {40}}(V(\Phi)+4e^\Phi)a^4
-{A\over {35}}(V(\Phi)-e^\Phi)a^5
\cr
&~~~~-{B\over {8800}}(16e^\Phi-11V(\Phi))(V(\Phi)+4e^\Phi)a^{10}+
{A\over {30030}}(75e^\Phi-39V(\Phi))(V(\Phi)-e^\Phi)a^{11}
\cr
&~~~~-{C\over {728}}\left({{27}\over 7}e^\Phi-{{45}\over {28}}\right)V(\Phi)
a^{12}+...
\cr}
\eqno(30)
$$
This wavefunction may be extrapolated to $a=a_b$ representing the boundary 
between the initial era and the inflationary epoch.  There it can be
matched with the no-boundary wavefunction
$$\Psi_{0NB}(a)={{Ai\left(K\left({{36}\over V}\right)^{2\over 3}\left(1-
{{a^2V}\over {6K}}\right)\right)}\over 
{Ai\left(K\left({{36}\over V}\right)^{2\over 3}\right)}}
\eqno(31)
$$
by equating the derivatives with respect to the scale factor up to second 
order.

A series solution of the form $\sum_n~c_n~(a-a_b)^n$, where $a_b$ is the
value of the scale factor at the boundary separating the initial Planck era
from the inflationary epoch, can be used to match the solution to the
third-order differential equation with the inflationary wave function.
A recursion relation is obtained for the coefficients of this series
$$\eqalign{{1\over 6}&\biggl[(n+5)(n+6)(n+7)c_{n+7}+2a_b(n+6)(n+7)(n+8)c_{n+8}
\cr
&~~~~~~~~~~~~~~~~~~~~~~~~~~~~~~~~~~~~~+a_b^2(n+7)(n+8)(n+9)c_{n+9}\biggr]
\cr
&+\biggl[(1-6a_b^5e^\Phi)(n+6)(n+7)c_{n+7}
+(a_b-(1+a_b^6e^\Phi))(n+7)(n+8)c_{n+8}
\cr
&~~~~~~~~-15a_b^4e^\Phi(n+5)(n+6)c_{n+6}-20a_b^3(n+4)(n+5)c_{n+5}e^\Phi
\cr
&~~-15a_b^2e^\Phi (n+3)(n+4)c_{n+4}-6a_b(n+2)(n+3)c_{n+3}e^\Phi
-(n+1)(n+2)c_{n+2} e^\Phi\biggr]
\cr
&+\biggl[(1+a_b^6e^\Phi-2a_b^6V(\Phi))(n+7)c_{n+7}+6a_b^5(e^\Phi-2V(\Phi))
(n+6)c_{n+6}
\cr
&~~~~~~~~+15a_b^4(e^\Phi-2V(\Phi))(n+5)c_{n+5}+20a_b^3(e^\Phi-2V (\Phi))
(n+4)c_{n+4}
\cr
&~~~~~~~~+15a_b^2(e^\Phi-2V(\Phi))(n+3)c_{n+3}+6a_b(e^\Phi-2V(\Phi))(n+2)
c_{n+2}
\cr
&~~~~~~~~~+(e^\Phi-2V(\Phi))(n+1)c_{n+1}\biggr]
\cr
&-3\biggl[c_{n+1}+5a_b c_{n+2}+10a_b^2 c_{n+3}+10 a_b^3 c_{n+4}
+5 a_b^4 c_{n+5}+a_b^5c_{n+6}\biggr]V(\Phi)~=~0
\cr}
\eqno(32)
$$
with $c_0$, $c_1$ and $c_2$ given by $\Psi_{inf.}(a_b)$, 
$\Psi_{inf.}^\prime(a_b)$ and ${1\over 2}\Psi_{inf.}^{\prime\prime}(a_b)$.  
These values may be determined, for example, from the no-boundary wavefunction,
whereas $a_b$, $e^\Phi$ and $V(\Phi)$ can be obtained from the 
time-dependence of the scale factor and the heterotic string potential. 

The large-$n$ limit of the recursion relation is 
$a_b^2c_{n+9}+2a_bc_{n+8}+c_{n+7}=0$ with solution 
$c_n\sim (-a_b)^{-n}n$ and substitution of this functional 
dependence into the series gives $\sum_n~(-a_b)^{-n}n~(a-a_b)^n={1\over 
{1+a_b^{-1}(a-a_b)}}={{a_b}\over a}$, which is the same divergent growth 
obtained earlier for the wave function as $a\to 0$.  Given standard values
for $e^\Phi$, $V(\Phi)$ and $a_b$, the recursion relation can
be solved numerically for the higher-order coefficients, and it may be 
confirmed that divergent behaviour is again obtained for the wavefunction 
near $a=0$.
 
Divergence at $a=0$ implies that a theory without dependence on the
scalar field could not be physically consistent unless a lower bound is
introduced for the scale factor.  The existence of a lower bound
would imply that the quantum theory must be defined on a minisuperspace 
$\{a(t),\Phi(t)~\vert~a(t)\ge a_0\}$ so that it describes perturbations about
classical solutions with spacelike sections of non-zero minimal radius.

\vskip 10pt
\noindent
{\bf 2. Solution to the Differential Equation for the Wavefunction in the 
Planck Era}

The change of variables $w=e^{-\Phi}$ leads to the following form for
the sixth-order differential equation:

$$\eqalign{
{{g_4^2}\over 6}&a^2w^2\left(a{{\partial^4\Psi}\over 
{\partial a^3 \partial w}}+6{{\partial^3\Psi}\over {\partial a^2 \partial w}}
\right)~+~Kw^3\left(w{{\partial^4\Psi}\over {\partial a^2 \partial w^2}}
+2{{\partial^3\Psi}\over {\partial a^2 \partial w}}\right)   
\cr
&=~a^4g_4^4\left[4a^3{{\partial \Psi}\over {\partial a}}+a^4 {{\partial^2\Psi}
\over {\partial a^2}}\right]~-~ag_4^2w^3\left(2{{\partial^3\Psi}\over
{\partial a \partial w^2}}+w{{\partial^4\Psi}\over {\partial a \partial w^3}}
\right)
\cr
&~~~~-{3\over 2}g_4^2 w^2\left(a{{\partial^2\Psi}\over {\partial a \partial w}}
+{{\partial \Psi}\over {\partial w}}+3w{{\partial^2\Psi}\over {\partial w^2}}
+w^2{{\partial^3\Psi}\over {\partial w^3}}\right)
\cr
&~~~~~~+{1\over {4a^6}}w^3\left(2{{\partial \Psi}\over {\partial w}}+46 w
{{\partial^2\Psi}\over {\partial w^2}}+115w^2 {{\partial^3\Psi}\over
{\partial w^3}}+75w^3{{\partial^4\Psi}\over {\partial w^4}}+16w^4{{\partial^5
\Psi}\over {\partial w^5}}+w^5{{\partial^6\Psi}\over {\partial w^6}}\right)
\cr}
\eqno(33)
$$
Given that the Mellin transform has the property ${\cal M}[f^{(n)};s]
=(-1)^n{{\Gamma(s)}\over {\Gamma(s-n)}}f^{*}(s-n)$, $\Psi^{*}(s,a)$
satisfies the equation
$$\eqalign{-a^2{{g_4^2}\over 6}&(s+1)\left[a {{d^3\Psi^{*}(s+1,a)}\over
{da}}+6{{d^2\Psi^{*}(s+1,a)}\over {da^2}}\right]+K(s+1)(s+2)
{{d^2\Psi^{*}(s+2,a)}\over {da^2}}
\cr
&~~=a^4g_4^4\left[4a^3 {{d\Psi^{*}(s,a)}\over {da}}+a^4{{d^2\Psi^{*}(s,a)}\over
{da^2}}\right]+ag_4^2(s+1)^2(s+2){{d\Psi^{*}(s+1,a)}\over {da}}
\cr
&~~~~~~~~+{3\over 2}g_4^2\left[a(s+1){{d\Psi^{*}(s+1,a)}\over {da}}+(s+1)^3
\Psi^{*}(s+1,a)\right]
\cr
&~~~~~~~~~+{1\over {4a^6}}[s^6+11s^5+50s^4+125s^3+205s^2+242s^s+152]
\Psi^{*}(s+2,a)
\cr}
\eqno(34)
$$
The difference-differential equation is
$$\eqalign{L_1(s,a)&E_1^2\Psi^{*}(s,a)+L_2(s,a)E_1\Psi^{*}(s,a)
+L_3(s,a)\Psi^{*}(s,a)=0
\cr
L_1(s,a)&=-K(s+1)(s+2)D_2^2+{1\over {4a^6}}[s^6+11s^5+50s^4+125s^3+205s^2+242s
+152]
\cr
L_2(s,a)&=a^2{{g_4^2}\over 6}(s+1)(aD_2^3+6D_2^2)+{3\over 2}g_4^2
(a(s+1)D_2+(s+1)^3)
\cr
&~~~~~~~~~~~~~~~~~~~~~~~~~~~~~~~~~~~~~~~~~~~~~~~~~~~~~~~~~~
+ag_4^2(s+1)^2(s+2)D_2
\cr
L_3(s,a)&=a^4g_4^4[4a^3D_2+a^4D_2^2]
\cr
E_1&\Psi^{*}(s,a)=\Psi^{*}(s+1,a)~~~~~~~~~D_2\Psi^{*}(s,a)={d\over {da}}
                                                             \Psi^{*}(s,a)
\cr}
\eqno(35)
$$
which has the asymptotic form
$${1\over {4a^6}}s^6u_{s+2}+ag_4^2(s+1)^2(s+2)D_2u_{s+1}+{3\over 2}g_4^2
(s+1)^2u_{s+1}\simeq 0
\eqno(36)
$$
with solution
$$\eqalign{{{u_{s+2}}\over {u_{s+1}}}~&\simeq~-{{(4a^6)}\over {s^3}}
                                                 \left[ag_4^2D_2+
                                                  {3\over 2}g_4^2\right]
\cr
u_{s+2}&\simeq (-1)^s{{(4a^6)^s}\over {\Gamma(s)^3}}
                                              \left[ag_4^2D_2+{3\over 2}g_4^2
                                                      \right]^s~u_2
\cr}
\eqno(37)
$$
The equation containing $u_2$ is
$$\eqalign{-KD_2^2u_2&+{1\over {4a^6}}152u_2+a{{g_4^2}\over 6}
(aD_2^3+6D_2^2)u_1+{3\over 2}g_4^2\left({5\over 3}aD_2+1\right)u_1
\cr
&~~~~~~~~~~~~~~~~~~~~+a^4g_4^4(4a^3D_2+a^4D_2^2)u_0=0
\cr}
\eqno(38)
$$
so that
$$\eqalign{u_2&=\left[KD_2^2-{{38}\over {a^6}}\right]^{-1}
\biggl\{ a{{g_4^2}\over 6}(aD_2^3+6D_2^2)u_1
+{3\over 2}g_4^2\left({5\over 3}aD_2+1\right)u_1
 +a^4g_4^4(4a^3D_2+a^4D_2^2)u_0 \biggr\}
\cr}
\eqno(39)
$$

For large $s$, the Mellin transform of the wavefunction is
$$\Psi^{*}(s,a)\approx (-1)^s{{(4a^6)^{s-2}}\over {\Gamma(s-2)^3}}
\left[ag_4^2D_2+{3\over 2}g_4^2\right]^{s-2}\left[KD_2^2-{{38}\over {a^6}}
\right]^{-1}\biggl\{a{{g_4^2}\over 6}(aD_2^3+6D_2^2)\Psi^{*}(1,a)
$$
$$\eqalign{
&~~~~~~~~~~~~~~~~~~~~~~~~~+{3\over 2}\left({5\over 3}aD_2+1\right)\Psi^{*}(1,a)
+a^4g_4^4(4a^3D_2+a^4D_2^2)\Psi^{*}(0,a)\biggr\}
\cr}
\eqno(40)
$$
Application of the inverse transform yields the wavefunction
$$\eqalign{\Psi(w,a)~&\approx~{1\over {2\pi i}}\int_{-i\infty}^{i\infty}
~(-1)^s{{(4a^6)^{s-2}}\over {\Gamma(s-2)^3}}
\left[ag_4^2D_2+{3\over 2}g_4^2\right]^{s-2}w^{-s}ds
\cdot\left[KD_2^2-{{38}\over {a^6}}\right]^{-1}
\cr
&~~~~~~~~~~~~~~~~~~~~\biggl\{a{{g_4^2}\over 6}(aD_2^3+6D_2^2)\Psi^{*}(1,a)
+{3\over 2}g_4^2\left({5\over 3}aD_2+1\right)\Psi^{*}(1,a)
\cr
&~~~~~~~~~~~~~~~~~~~~~~~~~~~~~~~~~~~~~~~~~~~~~~~~~~~
+a^4g_4^4(4a^3D_2+a^4D_2^2)\Psi^{*}(0,a)\biggr\}
\cr}
\eqno(41)
$$
After shifting the variable $s$ to $s-2$ and choosing the new contour
to be the imaginary axis, the integral has the form
$${{w^{-2}}\over {2\pi}}~\int_{-\infty}^\infty~\Gamma(-iy)^3~(-4a^6)^{iy}
\left[{{y~sinh~{\pi y}}\over \pi}\right]^3
\left[ag_4^2D_2+{3\over 2}g_4^2\right]^{iy}~w^{-iy}~dy
\eqno(42)
$$
or equivalently
$$\eqalign{{{w^{-2}}\over {16\pi^4}}&~\int_{-\infty}^\infty~\Gamma(-iy)^3~y^3~
e^{\left[3\pi+i\left[ln(-4a^6)+ln(ag_4^2D_2+{3\over 2}g_4^2)-ln~w)\right]
\right]y}
~dy
\cr
&-{{3w^{-2}}\over {16\pi^4}}
~\int_{-\infty}^\infty~\Gamma(-iy)^3~y^3~
e^{\left[\pi+i\left[ln(-4a^6)+ln(ag_4^2D_2+{3\over 2}g_4^2)-ln~w)\right]
\right]y}~dy
\cr
&+{{3w^{-2}}\over {16\pi^4}}
~\int_{-\infty}^\infty~\Gamma(-iy)^3~y^3~
e^{\left[-\pi+i\left[ln(-4a^6)+ln(ag_4^2D_2+{3\over 2}g_4^2)-ln~w)\right]
\right]y}~dy
\cr
&-{{w^{-2}}\over {16\pi^4}}
~\int_{-\infty}^\infty~\Gamma(-iy)^3~y^3~
e^{\left[-3\pi+i\left[ln(-4a^6)+ln(ag_4^2D_2+{3\over 2}g_4^2)-ln~w)\right]
\right]y}~dy
\cr}
\eqno(43)
$$

The function
$${1\over {\Gamma(s)}}=se^{\gamma s}~\prod_{n=1}^\infty\left(1+{s\over n}
                                                     \right)e^{-{s\over n}}
\eqno(44)
$$
has no poles or essential singularities in the finite plane, and it decays at 
a factorial rate if $Re~s>0$.  Consider the contour obtained by 
rotating the upper and lower halves of the imaginary axis towards the real 
axis.  Since the integrand vanishes on the circular contours $C_1$ and $C_2$, 
the integral over the imaginary axis equals the integral over a contour which
traverses the positive real axis in opposite directions and therefore
equals zero.

Given that equation (40) defines a close approximation to $\Psi^{\ast}(s,a)$
for $\vert s\vert\ge N_0$, a more accurate formula for the wavefunction is
$$\eqalign{\Psi(w,a)&\simeq{1\over {2\pi i}}\int_{-i\infty}^{i\infty}
~(-1)^s{{(4a^6)^{s-2}}\over 
{\Gamma(s-2)^3}}\left[ag_4^2D_2+{3\over 2}g_4^2\right]^{s-2}w^{-s}ds
\cdot\left[KD_2^2-{{38}\over {a^6}}\right]^{-1}
\cr
&~~~~~~~~~~~~~~~~~~~~\biggl\{a{{g_4^2}\over 6}(aD_2^3+6D_2^2)\Psi^{*}(1,a)
+{3\over 2}g_4^2\left({5\over 3}aD_2+1\right)\Psi^{*}(1,a)
\cr
&~~~~~~~~~~~~~~~~~~~~~~~~~~~~~~~~~~~~~~~~~~~~~~~~~~~
+a^4g_4^4(4a^3D_2+a^4D_2^2)\Psi^{*}(0,a)\biggr\}
\cr
&~~~+{1\over {2\pi i}}\int_{-iN_0}^{iN_0}~\Psi^{\ast}(s,a)w^{-s}ds
\cr
&~~~-{1\over {2\pi i}}\int_{-iN_0}^{iN_0}
~(-1)^s{{(4a^6)^{s-2}}\over 
{\Gamma(s-2)^3}}\left[ag_4^2D_2+{3\over 2}g_4^2\right]^{s-2}w^{-s}ds
\cdot\left[KD_2^2-{{38}\over {a^6}}\right]^{-1}
\cr
&~~~~~~~~~~~~~~~~~~~~\biggl\{a{{g_4^2}\over 6}(aD_2^3+6D_2^2)\Psi^{*}(1,a)
+{3\over 2}g_4^2\left({5\over 3}aD_2+1\right)\Psi^{*}(1,a)
\cr
&~~~~~~~~~~~~~~~~~~~~~~~~~~~~~~~~~~~~~~~~~~~~~~~~~~~
+a^4g_4^4(4a^3D_2+a^4D_2^2)\Psi^{*}(0,a)\biggr\}
\cr
&~=~{1\over {2\pi i}}\int_{-iN_0}^{iN_0}~\Psi^{\ast}(s,a)w^{-s}ds
\cr
&~~~-{1\over {2\pi i}}\int_{-iN_0}^{iN_0}
~(-1)^s{{(4a^6)^{s-2}}\over 
{\Gamma(s-2)^3}}\left[ag_4^2D_2+{3\over 2}g_4^2\right]^{s-2}w^{-s}ds
\cdot\left[KD_2^2-{{38}\over {a^6}}\right]^{-1}
\cr
&~~~~~~~~~~~~~~~~~~~~\biggl\{a{{g_4^2}\over 6}(aD_2^3+6D_2^2)\Psi^{*}(1,a)
+{3\over 2}g_4^2\left({5\over 3}aD_2+1\right)\Psi^{*}(1,a)
\cr
&~~~~~~~~~~~~~~~~~~~~~~~~~~~~~~~~~~~~~~~~~~~~~~~~~~~
+a^4g_4^4(4a^3D_2+a^4D_2^2)\Psi^{*}(0,a)\biggr\}
\cr}
\eqno(45)
$$
because the first integral vanishes.   The values of $\Psi^{\ast}(s,a)$
on the imaginary axis have not been given, but they can be defined by 
analytic continuation from the values on the real axis, which may be
deduced from $\{\Psi^{\ast}(s,a)\vert -2 < s < 2\}$ through the 
recursion relation.  The boundary conditions for the wavefunction can
be transformed to conditions on the Mellin transform.
$$\eqalign{\Psi^\ast(s,a_b)&=\int_0^\infty~w^{s-1}~\Psi(w,a_b)
=\int_0^\infty w^{s-1} \Psi_{inf.}(w,a_b)dw=\Psi^\ast(s,a_b)
\cr
{{\partial\Psi^\ast}\over {\partial a}}(s,a_b)&
                                           ={{\partial\Psi_{inf.}^\ast}\over 
                                                   {\partial a}}(s,a_b)
\cr
{{\partial^2\Psi^\ast}\over {\partial a^2}}(s,a_b)&=
           {{\partial^2\Psi_{inf.}^\ast}\over {\partial a^2}}(s,a_b)
\cr}
\eqno(46)
$$
Given $\Psi_{inf.}^\ast(s,a_b)$ for arbitrary values of $s$, the
the higher derivatives $\Psi^\ast(s,a_b),~-2<s<2$, can be obtained from the
difference-differential equation (34) and $\Psi^\ast(s,a),~-2<s<2$, can
be extrapolated to $a=0$.  

The $a\to 0$ limit of the integral (41) is given by
$$\eqalign{
lim_{a\to 0}&{{w^{-2}}\over {2\pi^4}}~\int_{-\infty}^\infty~\Gamma(-iy)^3
(-4a^6)^{iy}y^3~sinh^3\pi y~\left[ag_4^2D_2+{3\over 2}g_4^2\right]^{iy}
w^{-iy}dy 
\cr
&~~~~~~~~~~~~~~~\cdot\left[KD_2^2-{{38}\over {a^6}}\right]^{-1}
~\biggl\{a{{g_4^2}\over 6}(aD_2^3+6D_2^2)\Psi^{*}(1,a)
+{3\over 2}g_4^2\left({5\over 3}aD_2+1\right)\Psi^{*}(1,a)
\cr
&~~~~~~~~~~~~~~~~~~~~~~~~~~~~~~~~~~~~~~~~~~~~~~~~~~~~~~~~~~~~~~~~~~~~
+a^4g_4^4(4a^3D_2+a^4D_2^2)\Psi^{*}(0,a)\biggr\}
\cr
&~~~~~~\propto~{{w^{-2}}\over {2\pi^4}}\int_{-\infty}^\infty~\Gamma(-iy)^3
(-4)^{iy}\delta(y)y^3 sinh^3 \pi y~\left[{3\over 2}g_4^2\right]^{iy}w^{-iy}dy
\cr
&~~~~~~~~~~~~~~~~\cdot lim_{a\to 0}{{a^6}\over K}D_2^{-2}
~\biggl\{a{{g_4^2}\over 6}(aD_2^3+6D_2^2)\Psi^{*}(1,a)
+{3\over 2}g_4^2\left({5\over 3}aD_2+1\right)\Psi^{*}(1,a)
\cr
&~~~~~~~~~~~~~~~~~~~~~~~~~~~~~~~~~~~~~~~~~~~~~~~~~~~~~~~~~~~~~
+a^4g_4^4(4a^3D_2+a^4D_2^2)\Psi^{*}(0,a)\biggr\}
\cr}
\eqno(47)
$$
and again the integral over $y$ vanishes. 
Then
$$\eqalign{lim_{a\to 0}\Psi(w,a)~=~
{1\over {2\pi i}}&lim_{a\to 0}\int_{-iN_0}^{iN_0}\Psi^{\ast}(s,a)w^{-s}ds
\cr
-{1\over {2\pi i}}li&m_{a\to 0}\int_{-iN_0}^{iN_0}
(-1)^s{{(4a^6)^{s-2}}\over 
{\Gamma(s-2)^3}}\left[ag_4^2D_2+{3\over 2}g_4^2\right]^{s-2}w^{-s}ds
\cr
&~~~~~\left[KD_2^2-{{38}\over {a^6}}\right]^{-1}
\biggl\{g_4^2\left({a\over 6}(aD_2^3+6D_2^2)+
{3\over 2}\left({5\over 3}aD_2+1\right)\right)\Psi^{*}(1,a)
\cr
&~~~~~~~~~~~~~~~~~~~~~~~~~~~~~~~~~~~~~~~
+a^4g_4^4(4a^3D_2+a^4D_2^2)\Psi^{*}(0,a)\biggr\}
\cr
~=~{1\over {2\pi i}}&~lim_{a\to 0}~\int_{-iN_0}^{iN_0}~\Psi^{\ast}(s,a)w^{-s}ds
\cr}
\eqno(48)
$$
because the second integral vanishes by the regularity of the integrand in the 
$a\to 0$ limit as any divergences in $\Psi^{*}(0,a)$
and $\Psi^{*}(1,a)$ could be absorbed by the factor of $a^6$ multiplying the
operator $D_2^{-2}$.  This implies that the quantum theory would be consistent
in the $a\to 0$ limit when the scalar field degree of freedom is
included. 

Since the new terms in the sixth-order differential equation are given by
$$\eqalign{e^{-\Phi}P_\Phi e^{-\Phi}P_\Phi\biggl(g_4^2P_\Phi^{-1}e^\Phi a^4 P_a
&-{1\over {2a^3}}P_\Phi^2\biggr)(-a^3V(\Phi))\Psi
\cr
+~e^{-\Phi}P_\Phi e^{-\Phi}P_\Phi(-a^3V(\Phi))
\biggl(g_4^2P_\Phi^{-1}e^\Phi a^4& P_a~-{1\over {2a^3}}P_\Phi^2\biggr)\Psi
~+~e^{-\Phi}P_\Phi e^{-\Phi}P_\Phi a^6 V(\Phi)^2\Psi
\cr}
\eqno(49)
$$
regularity of the integrand in the expression for $\Psi(w,a)$ 
and therefore consistency of the quantum cosmology of the theory in the 
$a\to 0$ limit will be maintained in the presence of a potential V($\Phi$). 

\vskip 10pt
\centerline{\bf Acknowledgements}
\noindent
I would like to thank Professor Elmar Schrohe for his kind hospitality
at the Institut f{\"u}r Mathematik and useful conversations on differential
equation theory. I am grateful for the numerical programme of 
Dr W. Sch{\"o}bel to determine the coefficients of the series expansion 
solution of the third-order differential equation by recursive methods. 
This work has been supported by a Research Fellowship from 
the Alexander von Humboldt Foundation.
\vskip 15pt
\noindent
\centerline{\bf References}
\item{[1]} I. Antoniadis, J. Rizos, and K. Tamvakis, Nucl. Phys.
${\underline{B415}}$ (1994) 497-514
\item{[2]} S. Davis, Gen. Rel. Grav. ${\underline{32}}$(3) (2000) 541-551
\item{[3]} S. Davis and H. C. Luckock, Phys. Lett. ${\underline{459B}}$
(2000) 408-421
\item{[4]} H.-J. Schmidt, Phys. Rev. ${\underline{D49}}$(12)(1994) 6354 -6366 
\item{[5]} A. Vilenkin, Phys. Rev. ${\underline{D33}}$(12)(1986) 3560-3569
\item{[6]} X. Z. Zhong and D. Y. Liu, Acta Math. Appl. Sinica
${\underline{23}}$ (2000) 497-506
\item{[7]} A. D. Polyanin and V. F. Zaitsev, ${\underline{Handbook~of~
exact~solutions~for~ordinary}}$
\hfil\break
${\underline{differential~equations}}$ (Boca Raton: CRC Press, 1995)

\end